\documentclass[preprint,superscriptaddress]{revtex4-1}

\usepackage[T1]{fontenc}
\usepackage{amsmath}
\usepackage{amssymb}
\usepackage{graphicx}
\usepackage{bm}

\begin{document}

\title{Multidimensional topological strings by curved potentials: Simultaneous realization of mobility edge and topological protection}

 \author{Chun-Yan Lin}
\affiliation{Institute of Photonics Technologies, National Tsing Hua University, Hsinchu 300, Taiwan}

 \author{Giulia Marcucci}
 \affiliation{Department of Physics, University Sapienza, Piazzale Aldo Moro 2, 00185, Rome}

 \author{Gang Wang}
\affiliation{School of Physical Science and Technology, Soochow University, Suzhou 215006, China}

\author{You-Lin Chuang}
\affiliation{Physics Division, National Center for Theoretical Sciences, Hsinchu 30013, Taiwan}

 \author{Claudio Conti}
 \affiliation{Department of Physics, University Sapienza, Piazzale Aldo Moro 2, 00185, Rome}

 \author{Ray-Kuang Lee}
 \email{rklee@ee.nthu.edu.tw}
\affiliation{Institute of Photonics Technologies, National Tsing Hua University, Hsinchu 300, Taiwan}
\affiliation{Physics Division, National Center for Theoretical Sciences, Hsinchu 30013, Taiwan}

\begin{abstract}
By considering a cigar-shaped trapping potential elongated in a proper curvilinear coordinate, we discover a new form of wave localization which arises from the interplay of geometry and topological protection. The potential is modulated in its shape such that local curvature introduces a trapping potential. The curvature varies along the trap curvilinear axis encodes a topological Harper modulation. The varying geometry maps our system in a one-dimensional Andre-Aubry-Harper grating.
We show that a mobility edge exists and topologically protected states arises. These states are extremely robust with respect to disorder in shape of the string. The results may be relevant for localization phenomena in Bose-Einstein condensates, optical fibers and waveguides, and new laser devices, but also for fundamental studies on string theory. Taking into account that the one-dimensional modulation mimic the existence of a additional dimensions,
our system is the first example of physically realizable five-dimensional string.
\end{abstract}

\maketitle

Topological concepts pervade modern physics, including photonics, Bose-Einstein condensation (BEC), condensed matter physics, and high-energy physics. 
Topology enriches physical systems with non-trivial symmetries that protect specific features, as the energy eigenvalues of localized states, with respect to disorder. 
In addition, properly designed topological systems may link to analog additional dimensions in a way such that one can mimic multidimensional physics in the four-dimensional world~\cite{t1, t2, t3}.

Localized states may be induced at the interface between media with different topological features as Chern number. 
The onset of a topologically protected localized state is denoted as ``quantum-phase transition," and occurs as geometry is organized in a way such that symmetries change in different position. 
Edges state recently demonstrated to be very useful for a new generation of lasers, and attracted considerable attention ~\cite{e1, e2, e3, e4, e5, e6}

Beyond topology, geometrical features may also introduce different forms of localization, if a flat system is curved by a continuous deformation ~\cite{d1, d2, d3, d4, d5, d6}.
The local curvature creates an effective trapping potential. 
However, this effect is influenced by unwanted and random deformations.
For example, the interplay of disordered induced localization and geometry has been analyzed in~\cite{disorder}. 
Also curvature induced localization has recently been exploited to induce states at maximal curvature points in vertical cavity lasers~\cite{VCSEL}, as well as to speckle reduction in imaging
projection~\cite{speckle}.

\begin{figure}[hb]
\includegraphics[width=14cm]{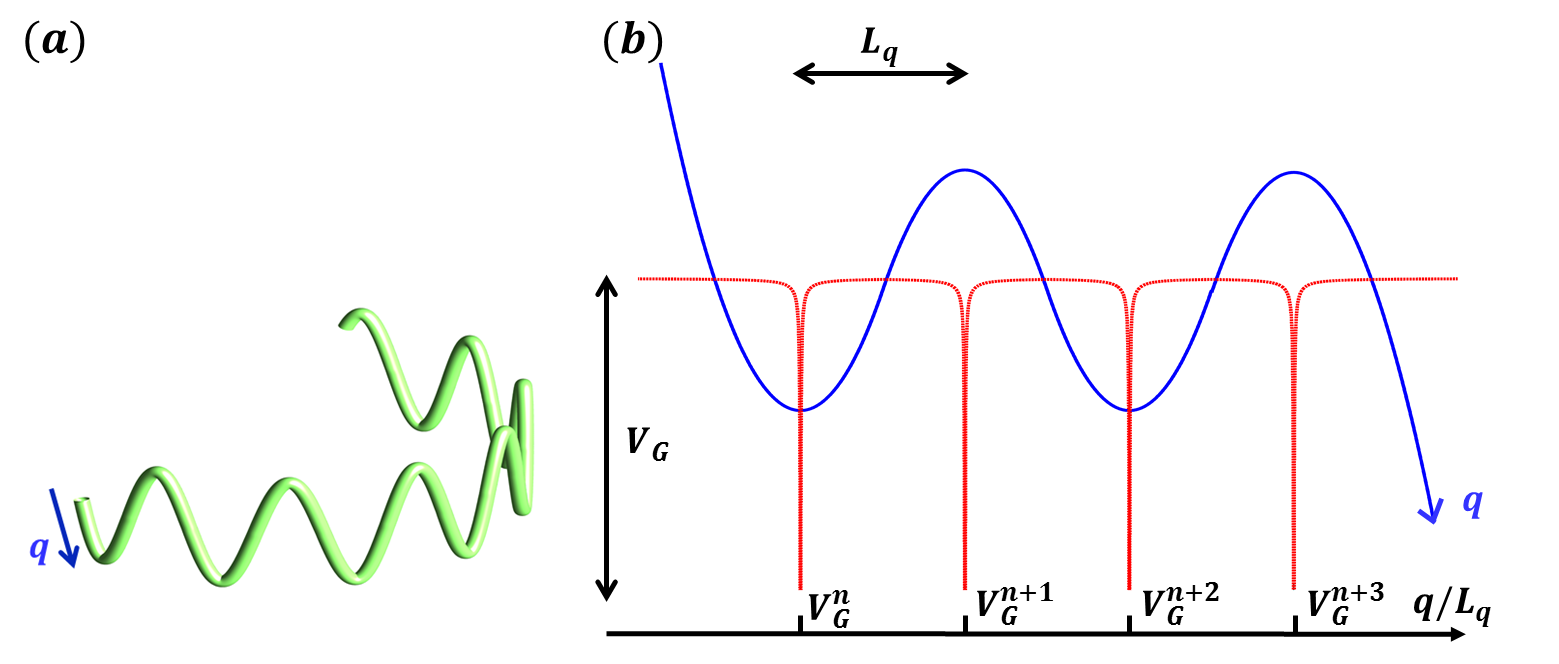}
\caption{(a) Sketch of a topological string made by a 3D trapping potential; (b) in the curvilinear coordinate $q$, we show four segments (blue line)
  with the corresponding potential $V_G$ (red line). $L_q$ is the distance between two parabolic segments.\label{Fig:1}}
\end{figure}
May we use topological protection to stabilizes bounds states due to local curvature? The connection and the interplay between the two forms of localization is unexplored. 
In addition, we know that some of the most important theoretical models of modern physics are two-dimensional strings embedded in multidimensional space. One can argue if one can realize these kind of systems in the laboratory.
Here we show that a strategy is given by our topological strings.
Their excitations, as localized states may represent analogues of fundamental physical particles~\cite{string}. Various systems are suitable for the experimental realization of these topological strings, such as BEC with elongated potential, curved optical waveguides and lasers, and polymer physics.

In this Letter we show that by combining geometrical and topological potentials, one obtains new kinds of trapped states.
We identify specific - open or closed - ``topological strings," such that the bound states of the curvature profile are robust with respect to deformation (see Fig.\ref{Fig:1}).
What we adopt as a reference model is the three-dimensional (3D) Schr{\"o}dinger equation
\begin{equation}
  i\,\Psi_t = -\nabla^2\Psi + V_{3D}\Psi.
  \end{equation}
  The concepts described in the following also apply in other systems, as lasers, optical fibers and waveguides.
  We consider a curved potential in a 3D space [Fig. 1 (a)], i.e., $V_{3D}=V_1(q_1)+V_\perp(q_2, q_3)$, where $q_1\equiv q$ is the longitudinal coordinate along the arc, and $q_2$, $q_3$ are the transverse coordinates ~\cite{d1, d6}.
Here, $V_\perp$ determines the transverse confinement along an arbitrary curve, whose curvilinear coordinate is given by $q$.
As long as $V_1(q_1)$ is a weak longitudinal trapping potential, it's effect becomes negligible for strong curvature, as detailed below. 
Specifically, we have a parabolic potential $V_\perp = w(q_2^2+q_3^2)$, $V_1(q_1)=w_1\, q_1^2$,  with $w \gg w_1$. 
Being $\lambda^2 = w/w_1$, the 1D reduction holds true as $\lambda^2 \rightarrow \infty$~\cite{d1}, letting $\Psi(q_1, q_2, q_3, t) = l_\perp \psi_\perp(q_2, q_3)\psi(q)\exp(-i\, E_\perp t)$, with $\int |\psi_\perp(q_2, q_3)|^2\, dq_2 dq_3 = 1$ and $E_\perp$ the transverse part of the eigenvalue. 
The transverse localization length is $l_\perp^{-2} = \int |\psi_\perp(q_2, q_3)|^4\, dq_2 dq_3$, and the normalization condition $\int |\psi(q)|^2\, dq = N l_\perp^{-2} \equiv P$.
For the parabolic potential above, the resulting 1D Schr{\"o}dinger equation with geometrical potential $V_G$ is given by
\begin{eqnarray}
  i\,\partial_t \psi = - \partial_q^2 \psi  + V(q)\psi,
  \label{eq:geo}
\end{eqnarray}
where $V = V_1 + V_G$, and $V_G$ is expressed in terms of the local curvature $K(q)$:
\begin{eqnarray}
V_G(q) = - \frac{K(q)^2}{4} = - \frac{k^2}{[1+4\, k^2\, x(q)^2]^3},
\end{eqnarray}
with $x(q)$ is given by the inverse of $4\,k\,q = 2\,k\,x \sqrt{1 + 4\,k^2\,x^2} + \sinh^{-1}(2\,k\,x)$. The maximum curvature is $2\,k = 1/R$, with $R$ denoting the radius of curvature. Figure~1(b) shows the local potential $V_G(q)$ for a four parabolic segments string.

We build a modulated potential by combining a series of parabolas as
in Fig.~\ref{Fig:1}(b).
Here, we consider a modulate path characterized by a repetition of parabolas merged with straight line.
By using Eq.~(\ref{eq:geo}) we aim at determining the path that realize a Harper modulation. The topological strings are 3D objects that are modulated in one dimension by a Harper potential, which is known to simulate additional synthetic  coordinates. Including also the time evolution, the system simulates a string in a five dimensional space.

\begin{figure}
\includegraphics[width=14cm]{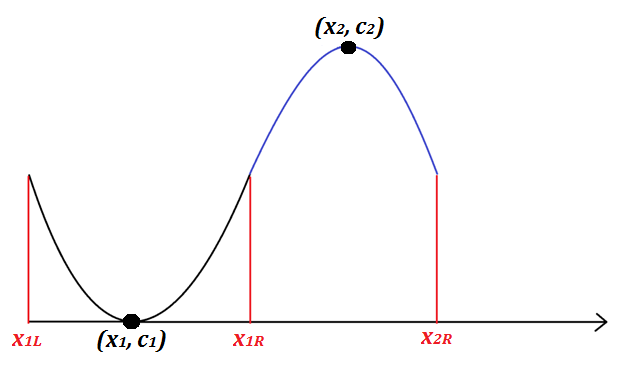}
\caption{The illustration to construct topological string. Here, we take $n = 1$ as an example. }
\end{figure}

The topological potential is a sequence of parabolas. Adopting a Cartesian set $(x, y)$, as in Fig. 1(b), each branch of the parabolas has extremum
(maximum or minimum) in $x_j$, $C_j$ with $C_j$ the local curvature is $k_j$.
In the Harper modulation, we calculate each portion of the parabola by fixing the period with the distance between two segments $L_q$ and the potential peak $V_G (n L_q) = V_0 + V_1\,\cos(2\pi \alpha n+\delta)$
to have a set of localized potentials. In the tight binding approximation, we have ($V_0$ is omitted hereafter without loss of generality)
\begin{eqnarray}
  t\,(\psi_{n+1}+\psi_{n-1}) + V_1\cos(2\pi\alpha n+\delta) \psi_n = E_n\, \psi_n.
  \label{Eq:t}
\end{eqnarray}
In Eq.~(\ref{Eq:t}) the hopping constant is denoted by $t$ and the
energy eigenvalue is $E_n$. 
In the Andre-Aubry-Harper (AAH) model,
the strength of the modulation is $V_1$, $\alpha$ is modulation frequency,
and  the phase term $\delta$ gives  the synthetic dimension~\cite{AAH}.
Compared to the well-known AAH model,  we describe our topological string by the length $L_q$ between two segments, and local curvature $k_n$.
For each branch of the parabola we have
\begin{eqnarray}
y=\pm k_n (x- x_n)^2 +c_n,
\end{eqnarray}
with the offset denoted as $c_n$.
The corresponding maximum of the potential is related to the curvature $V_G^n = -k_n^2$ .
By merging the different branches and imposing the continuity profile, we find that one can build a Harper modulation with a constant distance Lq between the maximum curvature points that are located at positions $x = x_n$ and $y = c_n$ with curvature coefficient $k^n$, with $n = 1, 2, \dots, N$.
To construct the topological strings, for each parabolic segments, in addition to the local curvature $k_n$ and the given distance $L_q$, we also have two  free parameters $x_n$ and $c_n$, corresponding to the center of the parabola and its offset.
Here, we fix the distance between two segment as a constant, i.e., $L_q = 6$, but vary the offset of the adjacent parabola to mimic the Harper modulation.
As for the curvature, the adjacent parabolas have the different sign, in order to have a smooth connection.
As illustrated in Fig. 2, to have a continuous derivative in the connected segment, we have $y_n^\prime (x_{nR}) = y_{n+1}^\prime (x_{nR} = x_{(n+1)L})$, or equivalently, $x_{n+1} = [(k_n+k_{n+1})x_{nR}-k_n x_n]/k_{n+1}$.
Then, for a fix value in $L_q$, we have $L_q = q_{nR}+q_{(n+1)L}$. 
For the offset, we have $c_{n+1} = c_n + k_n(x_{nR}-x_n)^2 \pm k_{n+1}(x_{nR}-x_{n+1})^2$, where we set the plus sign for the odd number in segment; and minus sign for the even number.

\begin{figure}
\includegraphics[width=14cm]{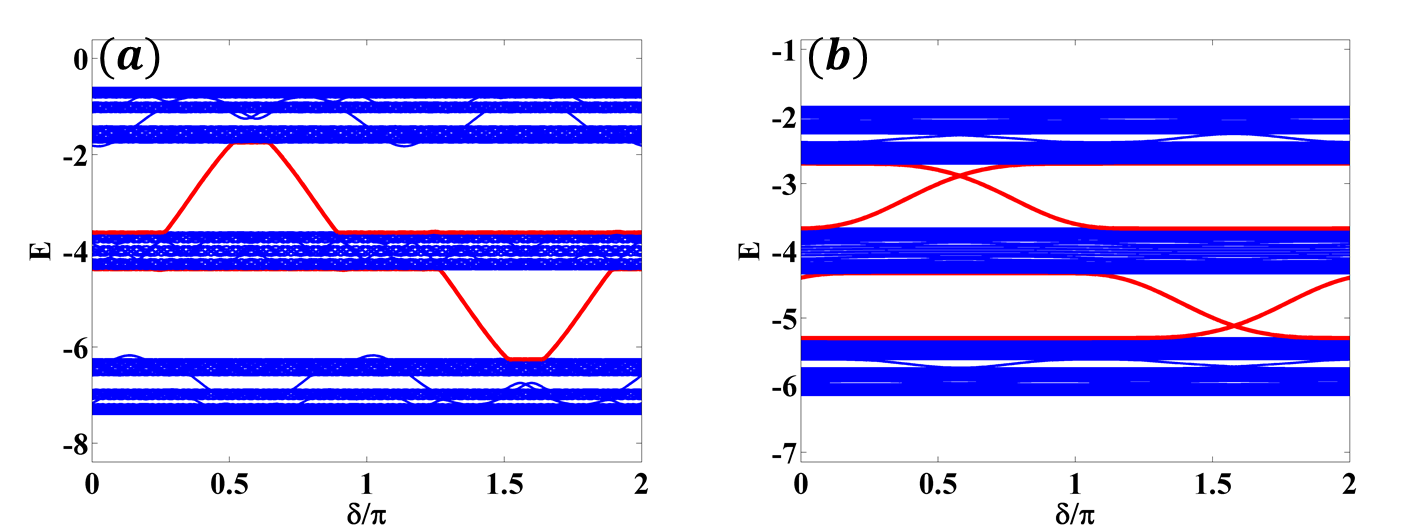}
\caption{Band diagrams for the AAH model in Eq.~(\ref{Eq:t}), for (a) $V_1/t = 3 > 2$ with all states localized; and (b) $V_1/t = 1 < 2$ with delocalized states. Red line gives the mobility edge states. Parameters are $V_0=-4$ and $\alpha = (1+\sqrt{5})/2$. \label{Fig:2}}\end{figure}

It is known that in AAH model there exists a critical value for the ratio between Harper modulation and the hopping constant, $V_1/t$. When $V_1/t > 2$, localized states are supported; for $V_1/t < 2$ all the modes are delocalized.
The transition from delocalized to localized modes is expected when $\alpha$ is an irrational number, for example $\alpha = (1+\sqrt{5})/2$. 
This case is referred to as the so called un-commensurate Harper model;
 it is specifically relevant as a one-dimensional model with a mobility edge in analogy with the Anderson model. 

 As illustrated in Fig.~\ref{Fig:2}, the band-diagrams for the AAH model
 in Eq.~(\ref{Eq:t}) support mobility edge states (red color),
 independently of $V_1/t$.
Instead of Brillouin zone defined in reciprocal space, here, the phase  term $\delta$ defined in the AAH model, see Eq.~(3), provides the synthetic dimension for  the band-diagram.
For the mobility edge states, the existence region in the band-gap is narrow when our system only supports localized states, see Fig. 3(a) for $V_1/t = 3$. Nevertheless, when delocalized states are supported, see Fig. 3(b) for $V_1/t = 1$, the existence region in the band-gap is broader.

To test the way the effective curvature potential mimic the Harper model, we study the mobility edge of the modes for $N = 20$ segments.
We then calculate the eigenstates of this potential and find a series of localized states and also delocalized functions for a different setting on $k_n \times L_q$. 

A key challenge is showing if the topological and geometrical potential $V(q)$ inherits the features of the discrete Harper modulation. 
In terms of the string parameters, the critical value for the localization-delocalization transition is defined by the product of the maximum local curvature and the length of string, i.e., $k_m \times L_q$. 
By numerical simulations, we have $k_m \times L_q \approx 8.44$. 

\begin{figure}
\includegraphics[width=14cm]{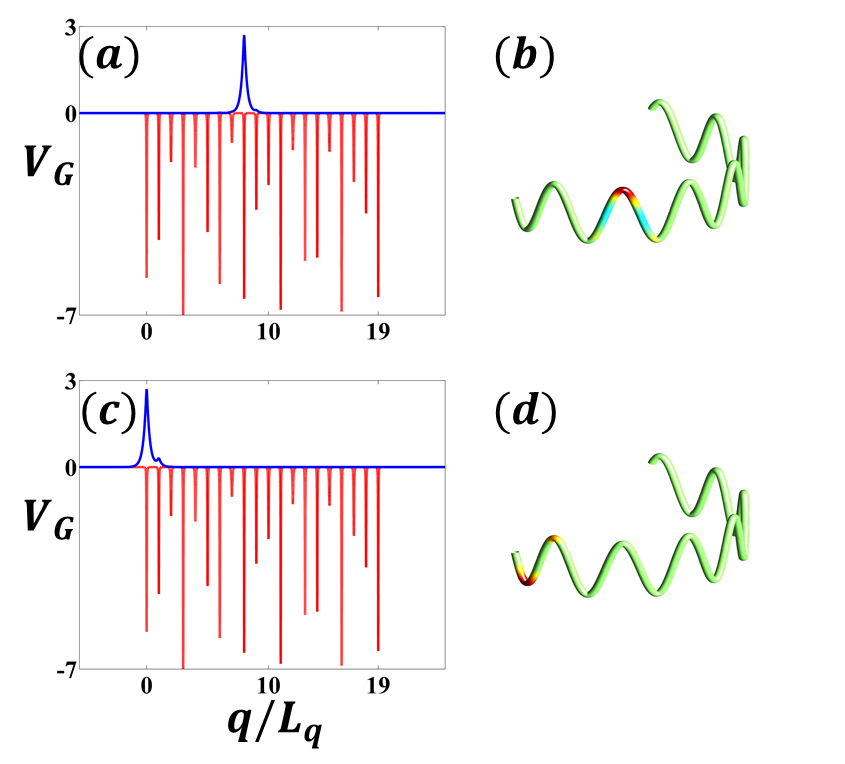}
\caption{(a) Localized states in the topological string:
  the blu line is the field profile, red line is the corresponding disordered potential $V_G$; (b) the corresponding topological string in 3D space. (c) Mobility edge states and  (d) the corresponding topological string in 3D space.
  Parameters: $V_0=-4$,  $V_1= 3.0$, $L_q = 6$ ($k_m \times L_q = 15.87 > 8.44$), and $\delta = 0.221$. }
\end{figure}

\begin{figure}
\includegraphics[width=14cm]{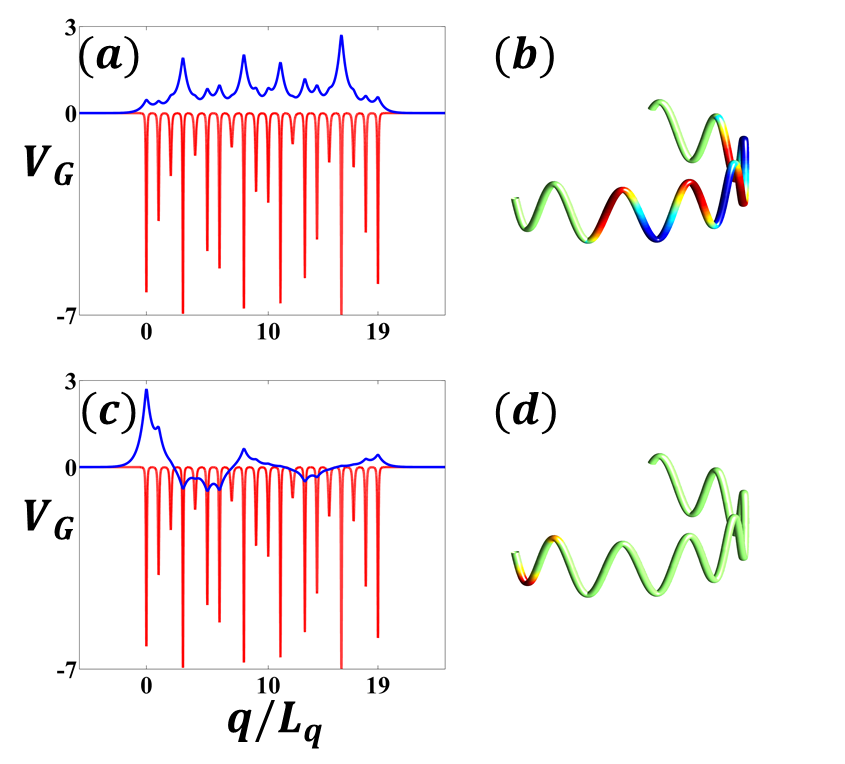}
\caption{(a) Delocalized states in the topological string, blue line is the field profile, and red line is the disordered potential $V_G$;
  (b) the corresponding topological string in 3D space. (c) Mobility edge states and  (d) the corresponding topological string in 3D space.
  Parameters: $V_0=-4$,  $V_1=3.0$, $L_q = 2.5$ ($k_m \times L_q = 6.61 < 8.44$), and $\delta = 0.221$.}
\end{figure}

For example, in Figs. 4 and 5, we show the potential profile for some representative cases, and the corresponding geometrical potential given by a series of localized traps with varying amplitudes $V_G$.
In Fig. 4(a-b), we illustrate the localized state supported in AAH model in Blue-colors, with the corresponding disorder potential $V_G$, based on the band-diagram illustrated in Fig. 3.
In addition to these localized states, in Fig. 4(c-d), one can see the  mobility edge states and  the corresponding topological string in 3D space. 

Figure 5(a-b) show the delocalized state supported in our topological string when we choose $k_m \times L_q = 6.61 < 8.44$. 
Even though the topological string do not support localized states, mobility edge states can still be found, see Fig. 5(c-d).
The supported mobility edge states are robust with respect to disorder in shape of the string.

\begin{figure}
\includegraphics[width=14cm]{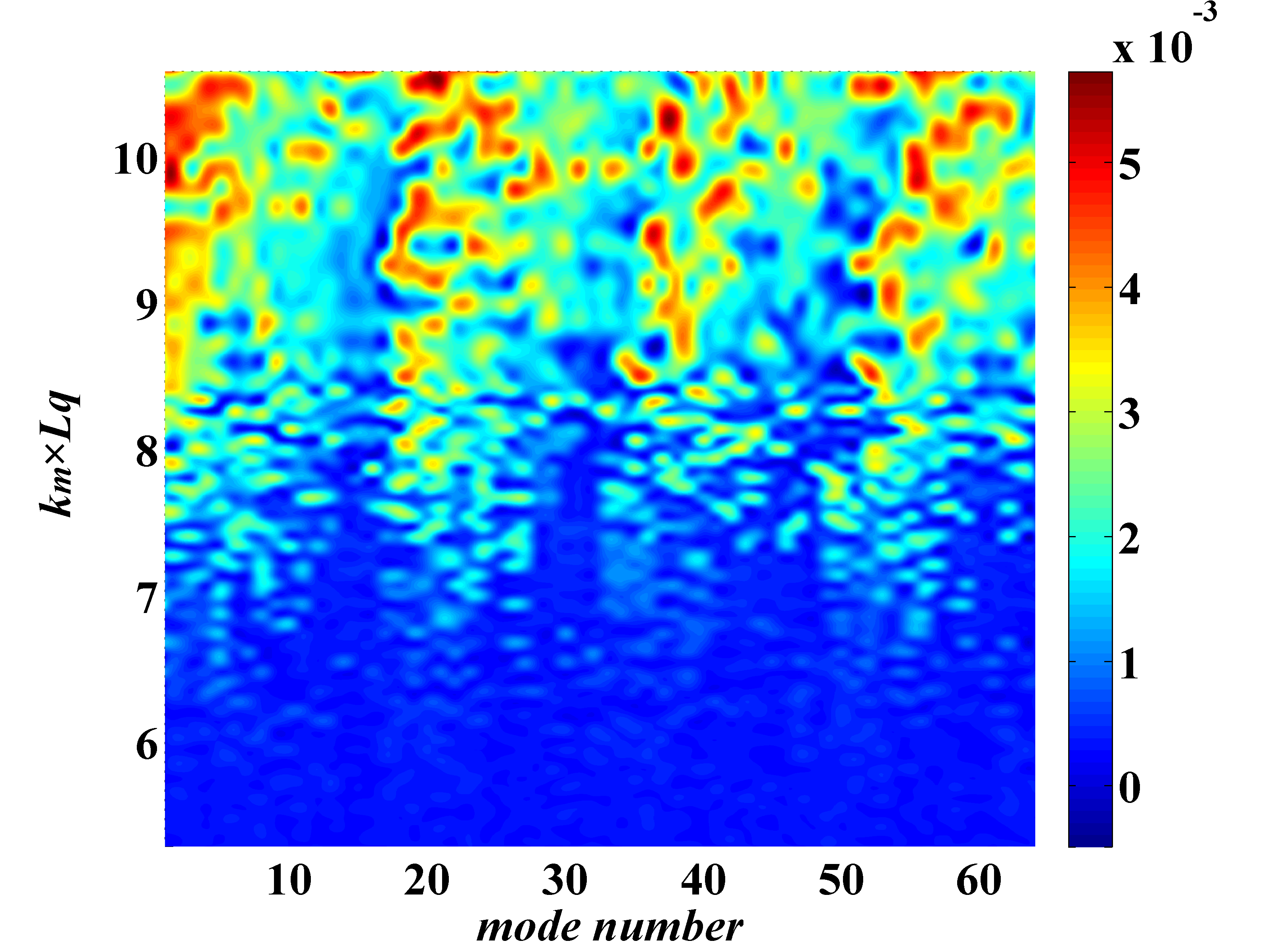}
\caption{The inverse participation ratio (IPR). One can see that there exists a transition in IPR at $k_m\times L_q\approx 8.44$. }
\end{figure}

To verify localization-delocalization transition in the topological strings, we show, in Fig. 6,  the inverse participation ratio (IPR)~\cite{IPR} below and above the critical ratio, defined by $k_m \times L_q$ for the description maximum local curvature and distance between two segments of a string.

In conclusion, we have predicted novel localized states
in string-like trapping potential with topological modulation.
Curvature and topological protection commit together to sustain
an entire new class of localized states, which are robust with
respect to fluctuations of system parameters and string geometry.
We have introduced for the first time this class of states,
which are expected to sustain novel and unexpected physical properties.
These states may have roles in novel information processing devices,
for classical and quantum computing, and for studying fundamental
physical process. It is remarkable that using topological string
one can experimentally test the resonances and features of excitations in
simulated multidimensional spaces, which can open the way to fundamental studies of string theory and related fields.

\section{acknowledgement}
The present research was supported by PRIN 2015 NEMO project (grant number 2015KEZNYM), H2020 QuantERA QUOMPLEX (grant number 731473), H2020 PhoQus (grant number 820392), PRIN 2017 PELM (grant number 20177PSCKT), Sapienza Ateneo (2016 and 2017 programs), and Ministry of Science and Technology of Taiwan (105-2628-M-007-003-MY4).

\end{document}